\documentclass[12pt]{article} 
\usepackage{graphicx}
\usepackage{amsbsy}

\title{Selective electron transfer between the quantum dots under the resonant pulse}
\author{Alexander V. Tsukanov and Leonid A. Openov$^{\dag}$}
\date{}

\begin{document}

\maketitle

\thanks{Institute of Physics and Technology, RAS, 117218, Moscow, Russia,
Nakhimovsky pr., 34, e-mail: tsukanov@ftian.oivta.ru,

$^{\dag}$Moscow Engineering Physics Institute (State University),
115409, Moscow, Russia, Kashirskoe sh., 31, e-mail:
opn@supercon.mephi.ru} \vskip 2mm

\vskip 4mm

\begin{abstract}
{\it The coherent quantum dynamics of an electron in the
quantum-dot ring structure under the resonant electromagnetic pulse
is studied theoretically. A possibility of the selective electron
transfer between any two dots is demonstrated. The transfer
probability as a function of the pulse and dot parameters is
calculated. It is shown that this probability can be close to
unity. The factors lowering the transfer probability in real
systems are
 discussed. The results obtained may be used in the
engineering of novel nanoelectronic devices for quantum bits
processing.}
\end{abstract}

\section{Introduction}

Rapid development of nanotechnology and unavoidable progress in
miniaturization of the basic elements of contemporaneous
microelectronics give rise to the new field of investigations, the
physics of low-dimensional structures and nanoelectronics. In
recent years, the great advances have been maid in the fabrication
of nanostructures and in the study of their properties [1],[2].
Much attention is now paid to the quantum dots (QD) - "artificial
atoms"[3] - which combine the properties of real atoms with the
properties imposed by the process of fabrication. A possibility to
use the QDs for quantum information processing is now extensively
discussed [4]-[7]. In these hypothetical devices, known as quantum
computers, the quantum information is encoded in ground and/or
excited orbital states of electron in the QD or in electron spin
degrees of freedom. In principle, it will be possible to realize
the quantum algorithms in such systems. To achieve this goal,
however, one has to overcome some challenges concerning with
initialization, processing, readout and storage of quantum
information. One of the main problems here is a coherent control of
evolution of electron states under the influence of external fields.

The behavior of low-dimensional objects is governed by quantum
effects. Coherent evolution of one-electron states in a double-dot
system upon the influence of a resonant laser pulse presents a good
example of such phenomena. As was recently shown [8],[9], the pulse
parameters (the frequency, duration, and amplitude) can be chosen
so as to drive the electron, localized initially in the ground
state in one of the dots, to the ground state of another dot via
the delocalized state, common for both dots and used as the
"transport state". If the states localized in different dots are
viewed as the Boolean states 0 and 1, the electron transfer
between them may be considered as the unitary operation NOT
[8]-[10].

Recently, an attempt has been made [11,12] to generalize the
results obtained for a double-dot system [8] to a chain-like multi-dot
system. It was shown that it is rather difficult to implement
the selective (addressed) electron
transfer between {\it arbitrary} two QDs.
This is because the probability amplitude to find an electron in a given
QD depends strongly on the location of this QD in a linear chain of
QDs with free boundary conditions.

In this work, we demonstrate the possibility of selective electron
transfer between arbitrary two QDs in the QD's ring structure under
a resonant laser pulse, the local bias voltages being applied to
those two dots. We derive an analytic expression for the transfer
probability which takes into account the possible deviations of
QD's and/or bias parameters from ideal ones as well as the detuning of the
laser pulse from the resonance.

\section{The model}

We consider a system composed of $N$ identical QDs arranged in a
ring, see Fig.1. We suppose that there are at least two
size-quantized levels in each QD. One of them, with the energy
${\varepsilon_{1}}$, corresponds to the ground state
$|1\rangle_{n}$ localized in a given QD with the number $n=1;...;N$. If the
value of $\varepsilon_{1}$ is close to the minimum of the potential
energy of an electron in the QD, and the height and/or width of the
potential barriers between QDs are large, then the ground state
wave functions of neighboring QDs overlap weakly because of the
strong localization of the wave functions $\langle
{\textbf{r}}|1\rangle_{n}$ in the corresponding QDs. In this case,
the ground state of the whole system may be considered as
$N$-fold degenerate with respect to the electron localization in
the QD system. We suppose that the excited levels $|2\rangle_{n}$
with the energy $\varepsilon_{2}$ (not necessarily second in the
level numeration) in each of the QDs lie close to the barrier edge.
In this case, the wave functions $\langle
{\textbf{r}}|2\rangle_{n}$ of the neighboring QDs overlap strongly,
resulting in the electron tunneling between QDs and splitting of
the excited levels into the subband of $N$ levels, each being
delocalized over the QD system. Since in the following we will use
the resonant (with respect to the external time-dependent field)
approximation, we neglect all levels
whose energies are far from $\varepsilon_{1}$ and $\varepsilon_{2}$.

The Hamiltonian of an excess electron added to the conduction band
(over the filled valence band) of such a QD structure, has the form
\begin{equation}
\hat{H_{0}}=\varepsilon_{1}\sum^{N}_{n=1}\hat{a}^{+}_{1,n}\hat{a}_{1,n}+\varepsilon_{2}\sum^{N}_{n=1}\hat{a}^{+}_{2,n}\hat{a}_{2,n}-V\sum^{N}_{n=1}(\hat{a}^{+}_{2,n+1}\hat{a}_{2,n}+h.c.),
\end{equation}
where $\hat{a}^{+}_{1,n}(\hat{a}_{1,n})$ and
$\hat{a}^{+}_{2,n}(\hat{a}_{2,n})$ are, respectively, the operators
of creation (annihilation) of an electron in the ground and excited
states of an isolated QD with the number $n$; $V$ is the matrix
element of electron tunneling between the excited states of
neighboring QDs. Note that $\hat{a}_{2,N+1}=\hat{a}_{2,1}$ since
the QD structure has the ring form. We don't show the spin indexes
explicitly in the Hamiltonian (1) since we consider a one-electron
problem.

Let an electron be initially localized in the ground state
$|1\rangle_{n_{1}}$ of a QD with the number $n_{1}$. In the absence
of an external field, the electron lifetime in the state
$|1\rangle_{n_{1}}$ is very long because of weak overlap of the
ground state wave functions of neighboring QDs. We assume this time
to be longer than all other characteristic times of the problem. We
wish to realize the selective electron transfer to the ground state
$|1\rangle_{n_{2}}$ of a QD with the number $n_{2}$, i. e., to
change the location of an electron in the QD system.

To realize the selective electron transfer between the QDs with the
numbers $n_{1}$ and $n_{2}$, we apply equal local bias voltages to
those QDs, thus shifting the energies of their ground and excited electron
states
($\varepsilon_{1}\rightarrow\varepsilon_{1}+\tilde{U},
\varepsilon_{2}\rightarrow\varepsilon_{2}+U$). The Hamiltonian
becomes

\begin{equation}
\hat{H}=\hat{H}_{0}+\tilde{U}(\hat{a}^{+}_{1,n_{1}}\hat{a}_{1,n_{1}}+\hat{a}^{+}_{1,n_{2}}\hat{a}_{1,n_{2}})+U(\hat{a}^{+}_{2,n_{1}}\hat{a}_{2,n_{1}}+\hat{a}^{+}_{2,n_{2}}\hat{a}_{2,n_{2}}),
\end{equation}
where, for the sake of simplicity, we ignore the changes in the
matrix elements of electron tunneling from the QDs with the numbers
$n_{1}$ and $n_{2}$ to neighboring QDs. This approximation is valid
if the local biases are small compared to $V$. For definiteness, we
will consider $\tilde{U}<0$ and $U<0$ (in general, $\tilde{U}\neq
U$, although they are of the same order). To
diagonalize the Hamiltonian (2), we go to operators
$\hat{a}^{+}_{k}=\sum^{N}_{n=1}C_{k,n}\hat{a}^{+}_{2,n}~
(k=1;...;N)$. We have
\begin{equation}
\hat{H}=\varepsilon_{1}\sum^{N}_{n=1}\hat{a}^{+}_{1,n}\hat{a}_{1,n}+\tilde{U}(\hat{a}^{+}_{1,n_{1}}\hat{a}_{1,n_{1}}+\hat{a}^{+}_{1,n_{2}}\hat{a}_{1,n_{2}})+\sum^{N}_{n=1}E_{k}\hat{a}^{+}_{k}\hat{a}_{k},
\end{equation}
where the energies $E_{k}$ of the delocalized levels and the
coefficients $C_{k,n}$ are determined by the set of equations
\begin{equation}
E_{k}C_{k,n}=\varepsilon_{1}C_{k,n}-V(C_{k,n-1}+C_{k,n+1})+UC_{k,n}(\delta_{nn_{1}}+\delta_{nn_{2}}),
~k=1;...;N,
\end{equation}
where $C_{k,N+1}=C_{k,1}$. The coefficients $C_{k,n}$ are the
probability amplitudes to find an electron occupying the $k$-th
state in the excited state $|2\rangle_{n}$ centered in a QD with
the number $n$. They satisfy the normalization condition
$\sum^{N}_{n=1}|C_{k,n}|^{2}=1$ for any $k=1;...;N.$

Expanding $C_{k,n}$ into a Fourier series,
$C_{k,n}=\frac{1}{N}\sum^{N}_{n=1}C_{k,n}\exp(i2\pi\ mn/N)$, we
obtain from Eq. (4) the following relationships between $C_{k,n_{1}}$
and $C_{k,n_{2}}$:
\begin{equation}
C_{k,n_{1}}=A_{k}C_{k,n_{1}}+B_{k}C_{k,n_{2}},~~
C_{k,n_{2}}=B^{*}_{k}C_{k,n_{1}}+A_{k}C_{k,n_{2}},
\end{equation}
where
\begin{equation}
A_{k}=-\frac{U}{N}\sum^{N}_{m=1}\frac{1}{E_{k}-\varepsilon_{2}+2V\cos(2\pi
m/N)}, B_{k}=-\frac{U}{N}\sum^{N}_{m=1}\frac{\exp(i2\pi
(n_{1}-n_{2})m/N)}{E_{k}-\varepsilon_{2}+2V\cos(2\pi m/N)}.
\end{equation}
It follows from Eqs. (5) and (6) that $|C_{k,n_{1}}|=|C_{k,n_{2}}|$ for
{\it any} delocalized level $k$ from the excited subband. This
 is important for the following consideration. Note
that the coefficients $C_{k,n_{1}}$ and $C_{k,n_{2}}$ may be chosen
real, so that $C_{k,n_{1}}=\pm C_{k,n_{2}}$.

Let the ac electric field $\mathbf{E}(t)=\mathbf{E_{0}}\cos(\Omega
t)$ be imposed on the QD system. The field frequency $\Omega$ is
close to the difference between the energy of one of the levels
from the excited subband, $E_{tr}$ (in the following called as the
"transport level"), and the ground state energies of an electron in
the QDs with the numbers $n_{1}$ and $n_{2}$ (hereafter we set the
Planck constant $\hbar=1$). In the resonant approximation [8,9], the
Hamiltonian reads
\begin{equation}
\begin{array}{l}
\hat H\left( t \right) = \sum\limits_{n = 1}^N {\left( {\varepsilon _1  + \tilde U(\delta _{nn_1 }  + \delta _{nn_2 } )} \right)\hat a_{1,n}^ +  \hat a_{1,n} }  + E_{tr} \hat a_{tr}^ +  \hat a_{tr}  - (e/m*c){\bf A}(t)\sum\limits_{n = 1}^N {\left( {{\bf p}_n \hat a_{tr}^ +  \hat a_{1,n}  + h.c} \right)}  =  \\
\sum\limits_{n = 1}^N {\left( {\varepsilon _1  + \tilde U(\delta _{nn_1 }  + \delta _{nn_2 } )} \right)\hat a_{1,n}^ +  \hat a_{1,n} }  + E_{tr} \hat a_{tr}^ +  \hat a_{tr}  - \left\{ {\frac{1}{2}\exp \left( { - i\Omega t} \right)\sum\limits_{n = 1}^N {\lambda _n \hat a_{tr}^ +  \hat a_{1,n} }  + h.c.} \right\}, \\
\end{array}
\end{equation}
where $\mathbf{p}_{n}=\langle tr|\hat{\mathbf{p}}|1\rangle_{n}$ are
the matrix elements of the momentum operator; $\mathbf{A}(t)$ is the
vector potential (we use the Lorenz gauge with zero scalar
potential and neglect the interaction term quadratic in the vector
potential), $m^{*}$ is the electron effective mass. In Eq. (7), we introduced
$\lambda_{n}=-\frac{ie}{m^{*}\Omega}\mathbf{E}_{0}\mathbf{p}_{n}$
making use of the well-known relationship between the vector potential and
the strength of an electric field with the frequency $\Omega$ and the
amplitude $\mathbf{E}_{0}$.

Here we point on a relationship between $\lambda_{n}$ and the
coefficients $C_{tr,n}$ in the expansion of the delocalized
transport state $|tr\rangle=\sum^{N}_{n=1}C_{tr,n}|2\rangle_{n}$ in
the states $|2\rangle_{n}$. From the definitions of $\lambda_{n}$
and $\mathbf{p}_{n}$ one obtains
\begin{equation}
\lambda_{n}=-\frac{ie}{m^{*}\Omega}\mathbf{E}_{0}\sum^{N}_{n'}C^{*}_{tr,n'}\cdot
_{n'}\langle2|\mathbf{\hat{p}}|1\rangle_{n}.
\end{equation}
Since the wave functions $\langle\mathbf{r}|2\rangle_{n}$ of the
excited states of QDs are centered in the vicinity of the
corresponding QDs, and the ground state wave functions are
localized in the QDs, one may suppose that ${}_{n'}\left\langle 2
\right|{\bf \hat p}\left| 1 \right\rangle _n = {}_n\left\langle 2
\right|{\bf \hat p}\left| 1 \right\rangle _n \delta _{nn'} $ , then
it follows from Eq. (8) that $\lambda_{n}=\lambda C_{tr,n}$, where
$\lambda=-\frac{ie}{m^{*}\Omega}\mathbf{E}_{0}\mathbf{p}$ and ${\bf
p} = {}_n\left\langle 2 \right|{\bf \hat p}\left| 1 \right\rangle
_n \,$.

We note that $\mathbf{p}\neq 0$ (i. e., $\lambda\neq 0$) only if a
certain relationship between the symmetries of the wave functions
$\langle\mathbf{r}|1\rangle_{n}$ and
$\langle\mathbf{r}|2\rangle_{n}$ takes place. For example,
$\mathbf{p}=0$ if both those functions have $s$-symmetry, while
$\mathbf{p}\neq0$ if one of them has $s$-symmetry and another has
$p$-symmetry. Besides, for the value of $\lambda_{n}$ to be
independent of $n$ (this is needed to meet the equality
$|\lambda_{n_{1}}|=|\lambda_{n_{2}}|$ which follows from the
condition $|C_{tr,n_{1}}|=|C_{tr,n_{2}}|$ obtained earlier and to
optimize the electron transfer between QDs), the vector
$\mathbf{p}$ (not only its absolute value) should be independent of
$n$. This is so if, e. g., the functions
$\langle\mathbf{r}|1\rangle_{n}$ have $s$-symmetry and the
functions $\langle\mathbf{r}|2\rangle_{n}$ have $p_{z}$-symmetry,
where $z$ axis is perpendicular to the QD ring plane, see Fig.1. In
this case, $\mathbf{E}_{0}$ should have non-zero component along
$z$ axis in order $\lambda\neq 0$.

In the resonant approximation, the evolution of the
electron wave function
\begin{equation}
|\Psi(t)\rangle=\sum^{N}_{n=1}B_{n}(t)|1\rangle_{n}\exp[-i(\varepsilon_{1}+\tilde{U}(\delta_{nn_{1}}+\delta_{nn_{2}}))t]+B_{tr}(t)|tr\rangle\exp(-iE_{tr}t)
\end{equation}
is governed by the non-stationary Schr\"{o}dinger equation
\begin{equation}
i\frac{\partial |\Psi(t)\rangle}{\partial
t}=\hat{H}(t)|\Psi(t)\rangle
\end{equation}
with the Hamiltonian (7) and the initial conditions
$B_{n}(0)=\delta_{nn_{1}}, B_{tr}(0)=0$. Our goal is to calculate
$B_{n}(t), B_{tr}(t)$ and thus to find $|\Psi(t)\rangle$. The
probability to find an electron in the ground state of the QD
with the number $n$ at a time $t$ is equal to
$p_{n}(t)=|B_{n}(t)|^{2}$.

We choose the lowest level of the excited subband as a transport
level. This choice is motivated by the following considerations.
First, since the excited states of the QDs are close to the barrier
edge, some of the upper levels of the excited subband may belong to
the continuum spectrum. In contrast, the energy of the lowest excited
level becomes smaller than $\varepsilon_{2}$, and hence the
corresponding wave function remains localized in the QD system
(although delocalized between different QDs). Second, the lowest
excited level at $U=0$ is non-degenerate for arbitrary $N$ and
remains non-degenerate at $U\neq 0$, while other excited levels at
$U=0$ constitute the set of the doubly degenerate states (except
for the highest level at even $N$). At $U\neq 0$, this degeneracy
is lifted, see Figs. 2 and 3, but the energy separations within the
doublets are small compared with the doublet separations themselves,
so that choosing any but the lowest of the excited delocalized levels
as a transport level makes it difficult to tune the laser to the resonance.

\section{Results and discussion}

We define the resonant frequency and the detuning from the
resonance as $\Omega_{r}=E_{tr}-(\varepsilon_{1}+\tilde{U})$ and
$\delta=\Omega-\Omega_{r}$, respectively. From Eqs. (7), (9), and
(10) we obtain the set of the differential equations for the
coefficients $B_n(t)$ and $B_{tr}(t)$:
\begin{equation}
\begin{array}{l}
\dot B_n (t) = i\frac{1}{2}\lambda _n^* B_{tr} (t)\exp \left[ {i\left( {\delta  - \tilde U(1 - \delta _{nn_1 }  - \delta _{nn_2 } )} \right)t} \right], \\
\dot B_{tr} (t) = i\frac{1}{2}\sum\limits_{n = 1}^N {\lambda _n B_n (t)\exp \left[ { - i\left( {\delta  - \tilde U(1 - \delta _{nn_1 }  - \delta _{nn_2 } )} \right)t} \right]}  \\
\end{array}
\end{equation}
where we took into
account that the states $|1\rangle_{n}$ and $|tr\rangle$ are the
eigenstates of the stationary Schr\"{o}dinger equation with the
eigenvalues
$\varepsilon_{1}+\tilde{U}(\delta_{nn_{1}}+\delta_{nn_{2}})$ and
$E_{tr}$, respectively.

Since the shift of the QD levels $\tilde{U}$ caused by the
local bias voltages applied to the QDs is finite and
detuning from the resonance is small (ideally, $\delta=0$),
one can take $|\delta|<<|\tilde{U}|$. Moreover, we assume
that the inequalities $|\delta|<<|\lambda|$ and
$|\lambda|<<|\tilde{U}|$ are satisfied, so that
$|\delta|<<|\lambda|<<|\tilde{U}|$. Then, as follows from
Eq. (11), the characteristic time $\sim 1/|\lambda|$ it takes for
the coefficients $B_{n_{1}}(t), B_{n_{2}}(t)$, and $B_{tr}$ to vary
is much longer than the corresponding time $\sim 1/|\tilde{U}|$ for the
coefficients $B_{n}(t)$ with $n\neq n_{1},n_{2}$,
and hence we have $|B_{n\neq n_{1},n_{2}}|\sim
(|\lambda|/|\tilde{U}|)|B_{n=n_{1},n_{2}}|<<|B_{n=n_{1},n_{2}}|$.
Thus, it is sufficient to restrict
ourselves with $n=n_{1}$ and $n=n_{2}$ in the sum (11):
\begin{equation}
\left\{ \begin{array}{l}
\dot{B}_{n_1 } \left( t \right) = i\frac{{\lambda _{n_1 }^* }}{2}B_{tr} \left( t \right) \cdot \exp \left( {i\delta t} \right), \\
\dot{B}_{n_2 } \left( t \right) = i\frac{{\lambda _{n_2 }^* }}{2}B_{tr} \left( t \right) \cdot \exp \left( {i\delta t} \right), \\
\dot{B}_{tr} \left( t \right) = i\frac{{\lambda _{n_1 } }}{2}B_{n_1 } \left( t \right) \cdot \exp \left( { - i\delta t} \right) + i\frac{{\lambda _{n_2 } }}{2}B_{n_2 } \left( t \right) \cdot \exp \left( { - i\delta t} \right), \\
\end{array} \right.
\end{equation}
\begin{equation}
\dot{B}_{n\neq
n_{1},n_{2}}(t)=i\frac{\lambda^{*}_{n}}{2}B_{tr}(t)\exp(i(\delta-\tilde{U})t).
\end{equation}
In this manner, the $(N+1)$-level problem can be
reduced to the $3$-level problem since there are only three
levels ($|1\rangle_{n_{1}}, |1\rangle_{n_{2}}$, and
$|tr\rangle$) relevant for the quantum dynamics of the system at
the resonance or close to it. This problem was recently solved
by us for the case $\delta\neq 0,
|\lambda_{n_{1}}|=|\lambda_{n_{2}}|$ in Ref. [8] and for the case
$\delta\neq 0, |\lambda_{n_{1}}|\neq|\lambda_{n_{2}}|$ in Ref. [9]. The
results obtained in Refs. [8,9] for the probabilities $p_{n}(t)$
can be directly applied to the
problem treated here. One can see [9] that at $\delta=0$, the
probability of electron transfer between QDs is
\begin{equation}
p_{n_{2}}(t)=\left(\frac{2|\lambda_{n_{1}}||\lambda_{n_{2}}|}{|\lambda_{n_{1}}|^{2}+|\lambda_{n_{2}}|^{2}}\right)^{2}\sin^{4}(\omega_{R}t),
\end{equation}
where
$\omega_{R}=\sqrt{|\lambda_{n_{1}}|^{2}+|\lambda_{n_{2}}|^{2}}/4$, and
hence the selective electron transfer between QDs takes place
in a time $T=\pi/2\omega_{R}$ if
$|\lambda_{n_{1}}|=|\lambda_{n_{2}}|$. A deviation of $\delta$
from zero and $|\lambda_{n_{1}}|$ from $|\lambda_{n_{2}}|$ causes
the value of $p_{n_{2}}(t)$ to decrease.

In this work, we account for a possible differences in the voltage biases
applied to two selected QDs. We take $U_{n_{1}}\neq U_{n_{2}}$
and $\tilde{U}_{n_{1}}\neq \tilde{U}_{n_{2}}$, then Eqs. (12) and(13) become
\begin{equation}
\left\{ \begin{array}{l}
\dot{B} _{n_1 } \left( t \right) = i\frac{{\lambda _{n_1 }^* }}{2}B_{tr} \left( t \right) \cdot \exp \left( {i\delta t} \right), \\
\dot{B} _{n_2 } \left( t \right) = i\frac{{\lambda _{n_2 }^* }}{2}B_{tr} \left( t \right) \cdot \exp \left( {i(\delta  - \Delta \varepsilon )t} \right), \\
\dot{B} _{tr} \left( t \right) = i\frac{{\lambda _{n_1 } }}{2}B_{n_1 } \left( t \right) \cdot \exp \left( { - i\delta t} \right) + i\frac{{\lambda _{n_2 } }}{2}B_{n_2 } \left( t \right) \cdot \exp \left( { - i(\delta  - \Delta \varepsilon )t} \right), \\
\end{array} \right.
\end{equation}
\begin{equation}
\dot{B}_{n\neq
n_{1},n_{2}}(t)=i\frac{\lambda^{*}_{n}}{2}B_{tr}(t)\exp(i(\delta-\tilde{U}_{n_{1}})t),
\end{equation}
where now $\Omega_{r}=E_{tr}-(\varepsilon_{1}+\tilde{U}_{n_{1}})$
and we designated $\Delta\epsilon=\tilde{U}_{n_{1}}-\tilde{U}_{n_{2}}$.

Using the following substitutions,
\begin{equation}
\left\{ \begin{array}{l}
B_{n_1 } \left( t \right) = \tilde B_{n_1 } \left( t \right) \cdot \exp \left( {i\frac{{\Delta \varepsilon }}{2}t} \right), \\
B_{n_2 } \left( t \right) = \tilde B_{n_2 } \left( t \right) \cdot \exp \left( { - i\frac{{\Delta \varepsilon }}{2}t} \right), \\
B_{tr} \left( t \right) = \tilde B_{tr} \left( t \right) \cdot \exp \left( { - i\left( {\delta  - \frac{{\Delta \varepsilon }}{2}} \right)t} \right), \\
\end{array} \right.
\end{equation}
we have from Eq. (15):
\begin{equation}
\left\{ \begin{array}{l}
\dot{\tilde B} _{n_1 } \left( t \right) + i\frac{{\Delta \varepsilon }}{2}\tilde B_{n_1 } (t) = i\frac{{\lambda _{n_1 }^* }}{2}\tilde B_{tr} \left( t \right), \\
\dot{\tilde B} _{n_2 } \left( t \right) - i\frac{{\Delta \varepsilon }}{2}\tilde B_{n_2 } (t) = i\frac{{\lambda _{n_2 }^* }}{2}\tilde B_{tr} \left( t \right), \\
\dot{\tilde B} _{tr} \left( t \right) - i\left( {\delta  - {{\Delta \varepsilon } \mathord{\left/
{\vphantom {{\Delta \varepsilon } 2}} \right.
\kern-\nulldelimiterspace} 2}} \right)\tilde B_{tr} \left( t \right) = i\frac{{\lambda _{n_1 } }}{2}\tilde B_{n_1 } \left( t \right) + i\frac{{\lambda _{n_2 } }}{2}\tilde B_{n_2 } \left( t \right). \\
\end{array} \right.
\end{equation}
Next, we express $\tilde B_{n_1 } (t)$ in terms of $\tilde B_{n_2}(t)$,
$\tilde B_{tr}(t)$, and $\dot{\tilde B}_{tr} \left( t\right)$
and $\tilde B_{tr} (t)$ in terms of $\tilde B_{n_2 }(t)$ and
$\dot{\tilde B}_{n_2 } \left( t \right)$ from the third and
second equations of the set (18), respectively. Inserting these expressions
into the first equation of the set (18), we obtain the equation for the coefficient
$\tilde B_{n_2 } (t)$:
\begin{equation}
 \begin{array}{l}
{\mathop {\tilde B}\limits ^{...}} _{n_2 } (t) - i\left( {\delta  -
{{\Delta \varepsilon } \mathord{\left/
{\vphantom {{\Delta \varepsilon } 2}} \right.
\kern-\nulldelimiterspace} 2}} \right){\mathop {\tilde B}\limits^{..}} _{n_2 } (t) + \left[ {\frac{{\left| {\lambda _{n_1 } } \right|^2  + \left| {\lambda _{n_2 } } \right|^2 }}{4} + \left( {\frac{{\Delta \varepsilon }}{2}} \right)^2 } \right]{\mathop {\tilde B}\limits^.} _{n_2 } (t) -  \\
- \frac{{\Delta \varepsilon }}{2}\left[ {\frac{{\left| {\lambda _{n_1 } } \right|^2  - \left| {\lambda _{n_2 } } \right|^2 }}{4} + \frac{{\left( {\delta  - {{\Delta \varepsilon } \mathord{\left/
{\vphantom {{\Delta \varepsilon } 2}} \right.
\kern-\nulldelimiterspace} 2}} \right)\Delta \varepsilon }}{2}} \right]{\mathop {\tilde B}\limits^{}} _{n_2 } (t) = 0. \\
\end{array}
\end{equation}
Taking into account that $B_{n_1 } (0) = 1$ and $B_{n_2 } (0) = B_{tr}
(0) = 0$, we have the following initial conditions: $\tilde B_{n_2 } (0)
= {\mathop {\tilde B}\limits^{.}}  _{n_2 } (0) = 0$, ${\mathop
{\tilde B}\limits^{ .. }} _{n_2 } (0) = - {{\lambda _{n_1 } \lambda
_{n_2 }^* } \mathord{\left/
{\vphantom {{\lambda _{n_1 } \lambda _{n_2 }^* } 4}} \right.
\kern-\nulldelimiterspace} 4}$.

The solution of Eq. (19) can be found exactly. It is, however, too combersome,
so here we restrict ourselves to the approximate solution for the case that
the deviations from the ideal case are small, i. e.,
${{\left| \delta  \right|} \mathord{\left/
{\vphantom {{\left| \delta  \right|} {\left| \lambda  \right|}}} \right.
\kern-\nulldelimiterspace} {\left| \lambda  \right|}},_{}^{} {{\left| {\Delta \varepsilon } \right|} \mathord{\left/
{\vphantom {{\left| {\Delta \varepsilon } \right|} {\left| \lambda  \right|}}} \right.
\kern-\nulldelimiterspace} {\left| \lambda  \right|}},_{}^{} {{\left| {\Delta \lambda } \right|} \mathord{\left/
{\vphantom {{\left| {\Delta \lambda } \right|} {\left| \lambda  \right|}}} \right.
\kern-\nulldelimiterspace} {\left| \lambda  \right|}} <  < 1$, where
$\Delta \lambda  = \left| {\lambda _{n_1 } } \right| - \left|
{\lambda _{n_2 } } \right|$:
\begin{equation}
\begin{array}{l}
B_{n_2 } \left( t \right) =  - \frac{{\lambda _{n_1 }^{} \lambda _{n_2 }^* }}{{\left| {\lambda _{n_1 } } \right|^2  + \left| {\lambda _{n_2 } } \right|^2  + \left( {\Delta \varepsilon } \right)^2 }}\exp \left( { - i\frac{{\Delta \varepsilon \left| {\lambda _{n_2 } } \right|^2 }}{{\left| {\lambda _{n_1 } } \right|^2  + \left| {\lambda _{n_2 } } \right|^2 }}t} \right) \cdot  \\
\cdot \left\{ {1 - \left[ {\cos \left( {2\omega _R t} \right) - i\frac{{\tilde \delta }}{{4\omega _R }}\sin \left( {2\omega _R t} \right)} \right]\exp \left( {i\frac{{\tilde \delta t}}{2}} \right)} \right\} \\
\end{array},
\end{equation}
where $\tilde \delta  = \delta  - \frac{{\Delta \varepsilon }}{2}
+ \frac{3}{2}\Delta \varepsilon \frac{{\left| {\lambda _{n_2 } }
\right|^2  - \left| {\lambda _{n_1 } } \right|^2 }}{{\left|
{\lambda _{n_1 } } \right|^2  + \left| {\lambda _{n_2 } }
\right|^2 }}$ and $\omega _R  = \frac{1}{4}\sqrt {\left| {\lambda
_{n_1 } } \right|^2  + \left| {\lambda _{n_2 } } \right|^2  +
\tilde \delta ^2  + \left( {\Delta \varepsilon } \right)^2 } $.
Then we have for the probability of the electron transfer:
\begin{equation}
\begin{array}{l}
p_{n_2 } \left( t \right) = \left| {B_{n_2 } \left( t \right)} \right|^2  = \left( {\frac{{2\left| {\lambda _{n_1 } } \right| \cdot \left| {\lambda _{n_2 } } \right|}}{{\left| {\lambda _{n_1 } } \right|^2  + \left| {\lambda _{n_2 } } \right|^2 +(\Delta\epsilon)^2}}} \right)^2  \cdot  \\
\cdot \left( {\sin ^4 (\omega _R t) + \sin ^2 \left( {\frac{{\tilde \delta t}}{4}} \right)\cos (2\omega _R t) + \frac{{\tilde \delta ^2 }}{{64\omega _R^2 }}\sin ^2 (2\omega _R t) - \frac{{\tilde \delta }}{{8\omega _R }}\sin \left( {\frac{{\tilde \delta t}}{2}} \right)\sin (2\omega _R t)} \right). \\
\end{array}
\end{equation}
For ${{\left| \delta  \right|} \mathord{\left/
{\vphantom {{\left| \delta  \right|} {\left| \lambda  \right|}}} \right.
\kern-\nulldelimiterspace} {\left| \lambda  \right|}},_{}^{} {{\left| {\Delta \varepsilon } \right|} \mathord{\left/
{\vphantom {{\left| {\Delta \varepsilon } \right|} {\left| \lambda  \right|}}} \right.
\kern-\nulldelimiterspace} {\left| \lambda  \right|}},_{}^{} {{\left| {\Delta \lambda } \right|} \mathord{\left/
{\vphantom {{\left| {\Delta \lambda } \right|} {\left| \lambda  \right|}}} \right.
\kern-\nulldelimiterspace} {\left| \lambda  \right|}} <  < 1$,
the maximum value of $p_{n_{2}}$ is
\begin{equation}
p_{n_2 }^{\max }  \approx 1 - {{\left[ {\left( {\Delta \lambda }
\right)^2  + \left( {\Delta \varepsilon } \right)^2  + \left( {\pi
^2 /8} \right)\left( {\delta  - {{\Delta \varepsilon }
\mathord{\left/
{\vphantom {{\Delta \varepsilon } 2}} \right.
\kern-\nulldelimiterspace} 2}} \right)^2 } \right]} \mathord{\left/
{\vphantom {{\left[ {\left( {\Delta \lambda } \right)^2  + \left( {\Delta \varepsilon } \right)^2  + \left( {\pi ^2 /8} \right)\left( {\delta  - {{\Delta \varepsilon } \mathord{\left/
{\vphantom {{\Delta \varepsilon } 2}} \right.
\kern-\nulldelimiterspace} 2}} \right)^2 } \right]} {\left| \lambda  \right|^2 }}} \right.
\kern-\nulldelimiterspace} {\left| \lambda  \right|^2 }}
\end{equation}
at $T=\pi/2\omega_{R}$, in agreement with Ref. [9].

If at $t=0$ an electron is placed in the superposition of states
$| 1 \rangle _{n_1 }$ and $| 1 \rangle _{n_2 }$, i. e.,
$B_{n_1 } (0) = \alpha $ and $B_{n_2 }(0) = \beta$
(so that $\left| \alpha  \right|^2  + \left| \beta
\right|^2  = 1$, i. e., $B_{n \ne n_1 ,n_2 } (0) = B_{tr} (0) = 0$),
then for the ideal structure ($\Delta \varepsilon  = \Delta \lambda=0$)
and at exact resonance $(\delta = 0)$ one obtains $B_{n_1 }
(T) =  - \beta $ and $B_{n_2 } (T) =  - \alpha $. Thus, if the
electron states $| 1 \rangle _{n_1 }$ and $| 1 \rangle _{n_2 }$
localized in different QDs are considered as the Boolean states
$|0\rangle$ and $|1\rangle$, respectively, so that their linear
combination is viewed as a quantum bit (qubit), then the resonance
pulse with duration $T$ is equivalent to the unitary quantum
operation NOT combined with the simultaneous change of the qubit phase
by $\pi$:
\begin{equation}
\hat U_{ideal} (T)\left| {\Psi (0)} \right\rangle  = \hat
U_{ideal} (T)\left( \begin{array}{l}
\alpha  \\
\beta  \\
\end{array} \right) = \left| {\Psi (T)} \right\rangle  = \left( \begin{array}{l}
\beta  \\
\alpha  \\
\end{array} \right)\exp (i\pi ).
\end{equation}
At non-zero values of $\delta ,_{}^{} \Delta \varepsilon$, and
$\Delta \lambda$, the fidelity of this operation, $F=$ $\left|
{\left\langle {{\hat U_{ideal} (T)\Psi (0)}} \mathrel{\left |
{\vphantom {{\hat U_{ideal} (T)\Psi (0)} {\hat U(T)\Psi (0)}}}
\right. \kern-\nulldelimiterspace} {{\hat U(T)\Psi (0)}}
\right\rangle } \right|^2 =$ $\left| { - \beta ^* B_{n_1 } (T) -
\alpha ^* B_{n_2 } (T)} \right|^2 $, differs from unity.

For the state-of-the art nanotechnology, it seems very difficult to
fabricate an ordered nanostructure composed of a great number of
nearly identical QDs. It is worth to mention another physical
system, for which the realization of the proposed scheme may be
possible, an array of phosphorous donor atoms embedded in a silicon
host [13], [14]. The modern techniques of the controlled
implantation of the phosphorous atoms into a silicon substrate [14]
allows, in principle, to fabricate the structures where the donors
serve as the centers of electron localization. Unlike an
"artificial" (QD) atoms, all "natural" atoms are identical, while
possible differences caused by implantation defects can be
minimized, by appropriate annealing. If all but one of $N$ donors
are ionized, the one-electron model studied in this paper may be
used to describe the evolution of an electron state.

Finally, since an electron resides in a solid rather than in a free space,
its unavoidable interactions with
other degrees of freedom can destroy
the unitary electron evolution under external
pulse. In particular, the processes of electron relaxation and dephasing
result to the decoherence.
This imposes some technological restrictions on the
structure and material parameters [15]. A detailed discussion of
decoherence effects in such structures will be presented elsewhere.

\section{Conclusions}
In this work, we have studied the unitary evolution of one-electron states
in an array of coupled QDs. We have shown that the selective electron
transfer between two arbitrary QDs under the influence of the resonant pulse
is possible. The probabilities to find an electron in the
ground states of those QDs as functions of time and the structure and pulse
parameters were calculated. We have shown that the probability of
electron transfer to the ground state of another QD may be very
close to unity. If we consider the states localized in two QDs as
logical variables $"0"$ and $"1"$, then the resonant pulse
applied to an arbitrary linear superposition of those states (qubit)
implements the quantum NOT operation with the simultaneous change of the
qubit phase by $\pi$.

We believe that since the selective electron transfer by virtue of
the "transport" state delocalized over all QDs may be performed as
a "one-step" operation, it has an advantage over the scheme based
on sequential "step-by-step" electron jumps between neighboring
QDs, as was recently suggested in Ref. [16]. The results obtained
in our work can be directly extended to other nano structures
composed of a large number of centers of electron localization, e.
g., the phosphorous donor in a silicon host, the atoms absorbed on
a surface, etc.

We are grateful to K.A. Valiev for his interest in this study and
to S.A. Dubovis who participated this work at its early stage.

\newpage

\includegraphics[width=\hsize]{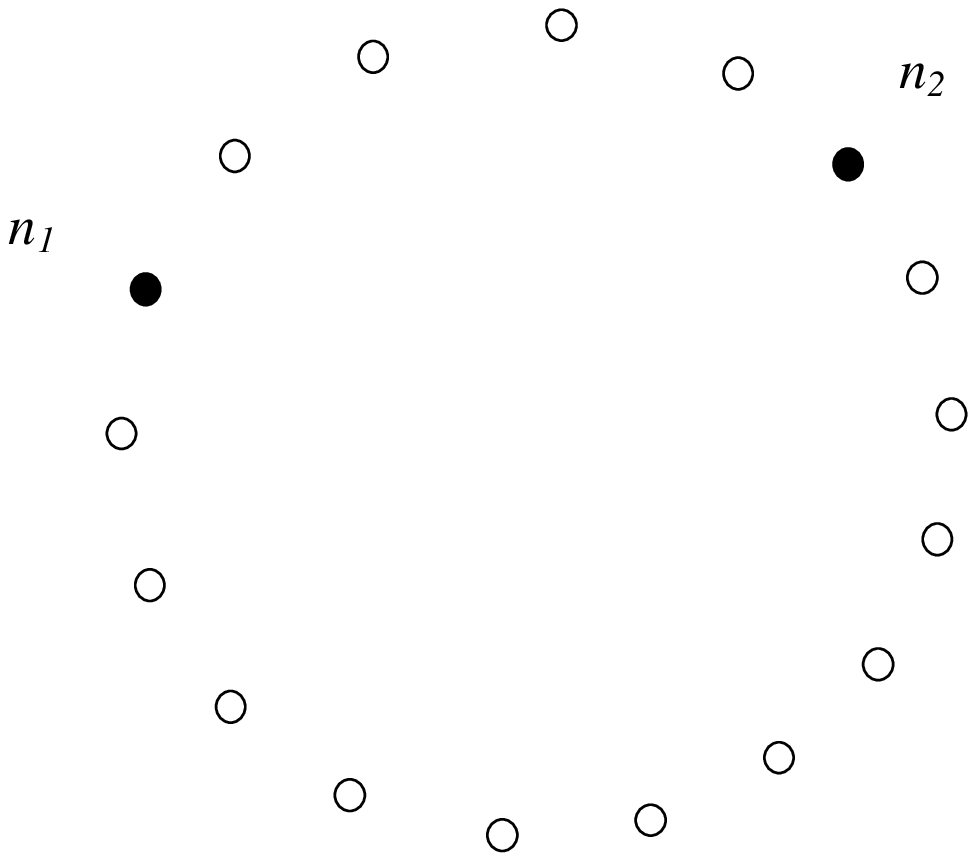}

\vskip 6mm

Fig. 1. The quasi-one-dimensional nanostructure
composed of identical QDs arranged in a ring. The
equal bias voltages are applied to the QDs with the numbers $n_{1}$
and $n_{2}$. The selective electron transfer between those QDs  is
performed under a resonant laser pulse.

\newpage

\includegraphics[width=\hsize]{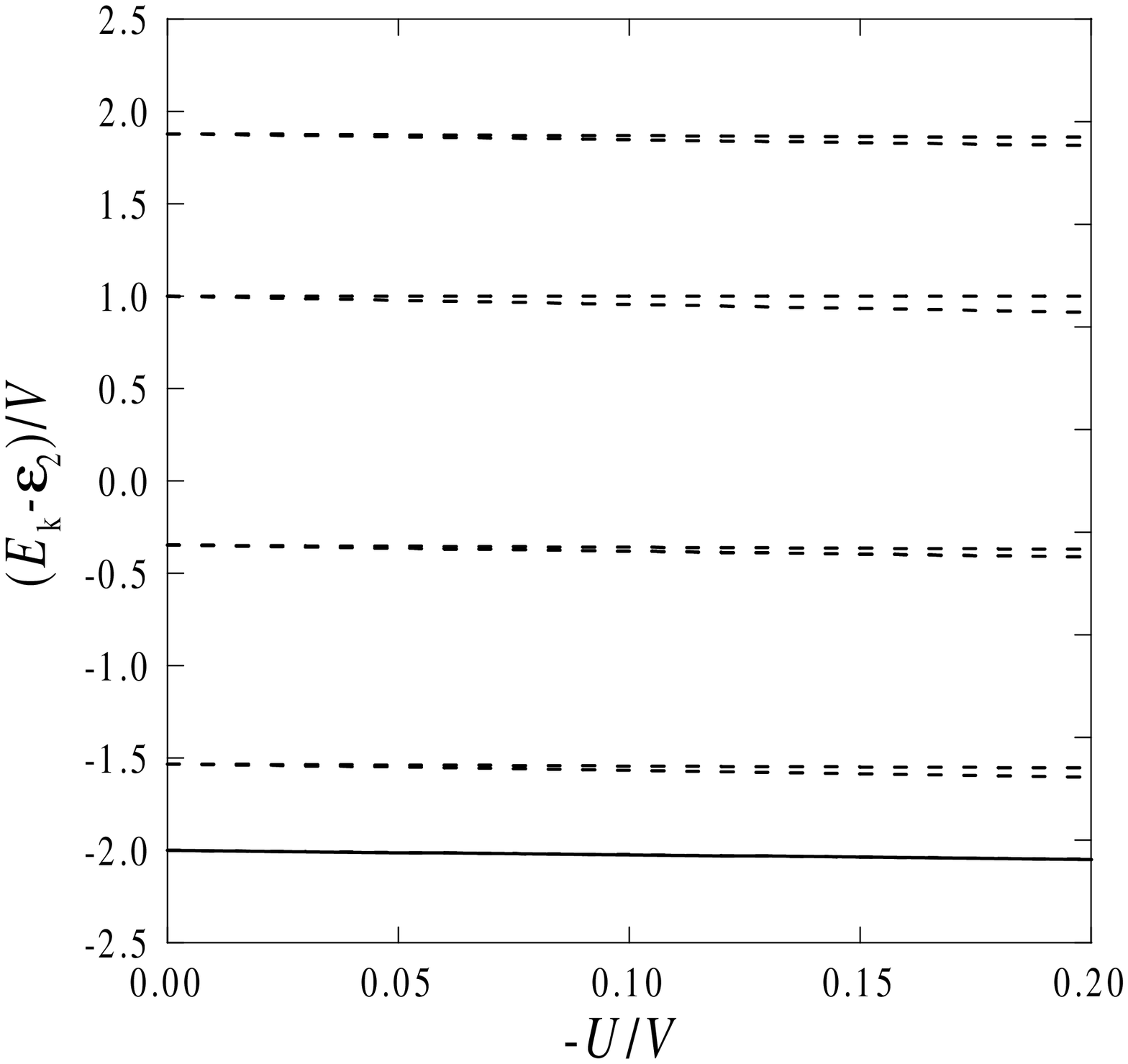}

\vskip 6mm

Fig. 2. The energy diagram of the $N$-dot structure shown in Fig. 1
shows the energies $E_{k}$ of the delocalized states in the excited subband
versus the
local bias voltages $U$ applied to the QDs with the numbers
$n_{1}$ and $n_{2}$.
$N=9,  n_{1}=1,  n_{2}=4$.
The solid and dash lines
correspond to the non-degenerate and degenerate at $U=0$ states,
respectively.

\newpage

\includegraphics[width=\hsize]{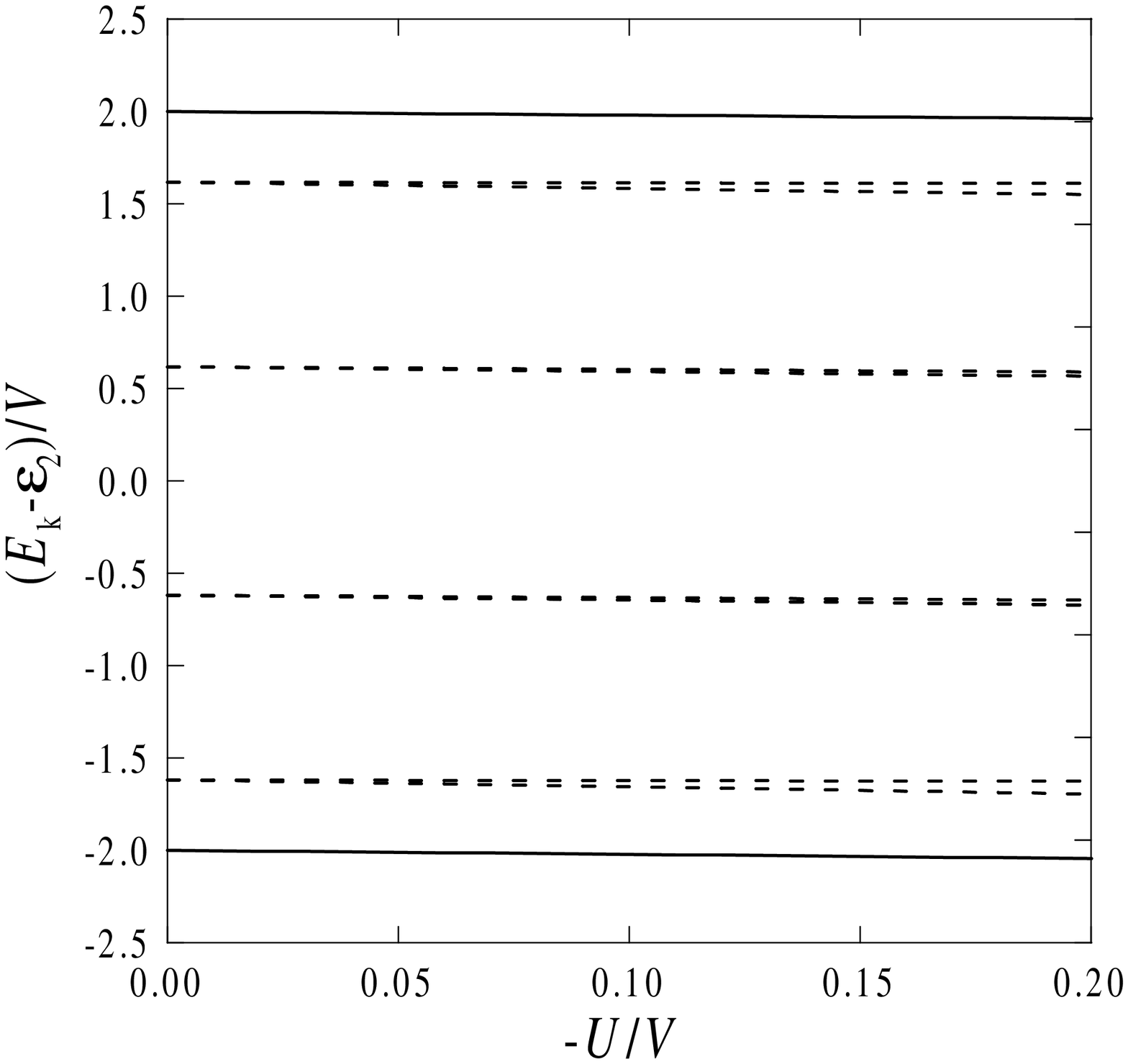}

\vskip 6mm

Fig. 3. The same as in Fig. 2 for $N=10,  n_{1}=1,  n_{2}=5$.

\end{document}